\documentclass[prl,aps,twocolumn,showpacs,amsmath,amssymb]{revtex4}
\usepackage{graphicx}% Include figure files
\usepackage{dcolumn}% Align table columns on decimal point
\usepackage{bm}% bold math

\begin{document}
\title{Novel dynamical effects and glassy response in strongly correlated electronic system }
\author{G. Y. Wang}
\author{ X. H. Chen}
\altaffiliation{Corresponding author} \email{chenxh@ustc.edu.cn}
\author{T. Wu, X. G. Luo, W. T. Zhang and G. Wu }
 \affiliation{Hefei
National Laboratory for Physical Science at Microscale and
Department of Physics, University of Science and Technology of
China, Hefei, Anhui 230026, People's Republic of China\\}

\date{\today}

\begin{abstract}
We find an unconventional nucleation of low temperature
paramagnetic metal (PMM) phase with monoclinic structure from the
matrix of high-temperature antiferromagnetic insulator (AFI) phase
with tetragonal structure in strongly correlated electronic system
$BaCo_{0.9}Ni_{0.1}S_{1.97}$. Such unconventional nucleation leads
to a decease in resistivity by several orders with relaxation at a
fixed temperature without external perturbation. The novel
dynamical process could arise from the competition of strain
fields, Coulomb interactions, magnetic correlations and disorders.
Such competition may frustrate the nucleation, giving rise to a
slow, nonexponential relaxation and "physical aging" behavior.
\end{abstract}

\pacs{71.27.+a; 71.30.+h; 72.90.+y}

\maketitle
\newpage
Slow, nonexponential relaxations of glassy dynamics have been
widely observed in doped semiconductors\cite{monroe}, strongly
disordered indium-oxide films\cite{ovadyahu}, and various granular
metals\cite{martinez,bielejec}. Such nonergodic behavior is very
interesting because one normally expects electron systems to relax
rather rapidly. Many glassy systems exhibit a nonstationary
behavior that has been described as "physical aging"\cite{hodge}.
Recently, the electron glass has received renewed interest
\cite{orignac,pastor} as the subject of electron-electron
interactions has become a central topic in understanding the
metal-insulator transition (MIT) in two dimensions\cite{simonian}.
Such glass behavior is believed to be associated with the
interplay between disorder and strong electronic
correlations\cite{vaknin}. Phase separation widely observed in
strongly correlated electronic system\cite{tranquada,uehara}
provides the possibility  for appearance of locally metastable
states, giving rise to the self-organized inhomogeneities
(disorders). Phase separation (PS) scenario appears as
particularly favorable for the existence of out-of-equilibrium
features.  Therefore, it is expected that the glass behavior
occurs in the strongly correlated electronic system due to
interplay between disorder and strong electronic
correlations\cite{vaknin}, especially in phase separation region.
Indeed, a pronounced glassy response\cite{eblen} and a memory
effect have been recently observed in phase separated
manganites\cite{levy}. The interplay of strong electronic
correlations and disorder is believed to be responsible for many
new phenomena occurring in complex materials in the MIT
region\cite{miranda}. Therefore, Novel findings in complex
materials with strongly electronic correlations in the MIT region
clearly deserves further study.

$BaCoS_2$ is a Mott-Hubbard insulator having $Co_2S_2$ layers with
spin-1/2 Co ions that order antiferomagnetically at 310 K, and
properties shared by members of the high $T_c$ cuprates. Such
quasi-two dimensional system $BaCoS_2$ with $CoS$ conducting
planes separated by insulating BaS rocksalt sheets is structurally
analogous to the high-$T_c$ cuprates\cite{martinson}. Substitution
of Ni for Co leads to a first-order transition from an
antiferromagnetic insulator (AFI) to a paramagnetic metal (PMM)
upon cooling at $T_c$ in the layered $BaCo_{0.9}Ni_{0.1}S_{2-y}$
system for $0.05 \leq y \leq 0.20$\cite{martinson}. Such AFI-PMM
phase transition is associated with a structural change from
high-temperature tetragonal (HTT) phase to low-temperature
monoclinic (LTM) phase\cite{schweitzer}.

In this letter, we report a novel dynamical process for the MIT
transition due to the interplay of strong electronic correlations
and disorder in $BaCo_{0.9}Ni_{0.1}S_{1.97}$ close to the
composition with MIT. In novel dynamical process, the competition
of strain fields, Coulomb interactions, magnetic correlations and
disorder in this correlated electronic system frustrates the
nucleation, giving rise to a slow, nonexponential relaxation of
glassy dynamics and "physical aging" behavior.

Figure 1 shows temperature dependence of resistivity measured in
two different sequences of cooling and warming cycles with
starting temperature of 400K and 300K for the sample
$BaCo_{0.9}Ni_{0.1}S_{1.97}$, respectively. In the first round of
the first sequence, resistivity was measured with cooling the
sample from 400 to 5 K, subsequently warmed to 400K. The sample
shows an insulating behavior with a kink at $T_{s}\sim 65K$ upon
cooling. Such kink, referred to a first-order
\emph{AFI-$I^\prime$} transition, has been observed in sample
$BaCo_{0.9}Ni_{0.1}S_{1.9}$ after complete suppression of the
AFI-PMM transition by pressure\cite{looney}. A hysteresis is
observed between curves recorded upon cooling and subsequent
heating. Continuously, resistivity was measured in the second
round of 400K $\rightarrow$ 5K $\rightarrow$ 300K. The resistivity
obtained on cooling in the second round is nearly the same as that
in first round, but the hysteresis becomes larger. The sample was
continuously measured in the third round from 300K $\rightarrow$
5K $\rightarrow220K$. Upon cooling, the resistivity shows the
similar behavior to the second round, but the resistivity below
\begin{figure}[t]
\includegraphics[width=9cm]{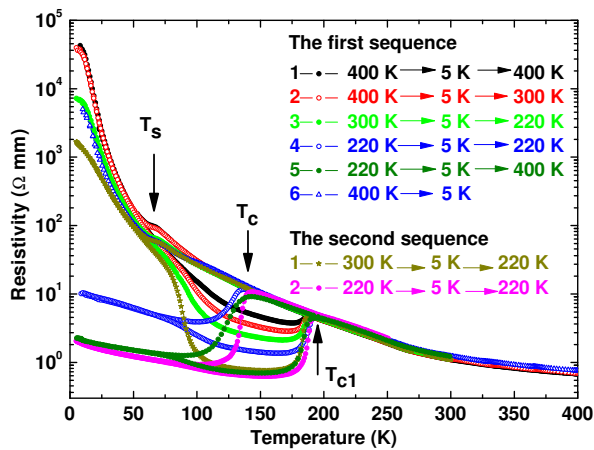}
\caption{Temperature dependence of resistivity measured with
different cooling and warming cycles in the two sequences in
$BaCo_{0.9}Ni_{0.1}S_{1.97}$ close to the phase boundary between
an AFI at lower sulphur deficiency and a PMM at higher sulphur
deficiency. Through all measuurements, the rate of cooling and
warming was kept constant (3 K/min).\\ }
\end{figure}120 K is less than that in the first and second
round with much larger hysteresis. Striking feature is that a
first-order phase transition at $T_c \sim 140 K$ from insulator to
metal with hysteresis occurs in the fourth round of $220 K
\rightarrow 5 K\rightarrow 220K$. It is similar to the AFI-PMM
transition observed in the sample with large sulphur
deficiency\cite{martinson}. In the fifth round from $220 K
\rightarrow 5 K\rightarrow 400K$, resistivity shows similar
behavior to that in the fourth round except that the transition is
sharper with a larger reduction of resistivity. These results
indicate that resistivity is strongly dependent on cooling and
warming cycles. The hysteresis becomes larger with cooling and
warming cycles, consequently the first-order phase transition from
AFI with HTT structure to PMM with LTM structure is induced. It
indicates that an irreversible and memory behavior occurs in
cooling and warming cycles. After the fifth round, another
intriguing behavior is observed with cooling the sample from
\emph{400 K to 5 K}, the first-order phase transition from AFI
with HTT structure to PMM with LTM structure is gone and the
resistivity re-exhibits an insulating behavior. Disappearance of
the phase transition with cooling sample from \emph{400 K
}suggests that the temperature of \emph{400 K} can remove the
memory effect, so that effect of cooling and warming cycle on
resistivity is gone.

In order to further understand the effect of \emph{400 K} on
resistivity, resistivity was measured in the second sequence with
starting temperature of 300 K after keeping the sample at room
temperature for more than ten days. As shown in Fig.1, the
resistivity measured with cooling sample from \emph{300 K} in the
second sequence exhibits an insulating behavior with the
\emph{AFI-$I^\prime$} transition, being similar to that observed
in the first sequence. But resistivity obtained with cooling
sample from \emph{300 K} is less than that from \emph{400 K} below
120 K (much less than one order of magnitude at 5 K). The
hysteresis obtained in the first round of second sequence is much
larger than that in the first round of the first sequence.
Continuously, resistivity was measured in the second round of $220
K \rightarrow 5 K\rightarrow 220K$, a first-order phase transition
similar to that observed in the first sequence takes place at
$\sim 140 K$. The common feature shared in the two sequences is
that the phase transition can be only observed after the cooling
and warming cycle of $300 K \rightarrow 5 K\rightarrow 220K$. As
shown in Fig.1, the resistivity at 5 K obtained with the different
cooling and warming experience of the sample can change by more
than four orders of magnitude. It suggests that an insulating
phase and a metallic phase coexist in the sample at low
temperature, and a \emph{nucleation} of the metallic phase from
the insulating matrix occurs with cooling and warming cycles.
Therefore, a phase separation ocurs in
$BaCo_{0.9}Ni_{0.1}S_{1.97}$ as observed in manganites\cite{levy}.

\begin{figure}[b]
\includegraphics[width=9cm]{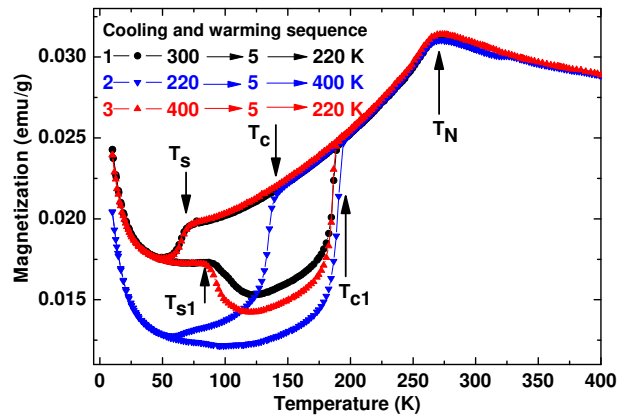}
\caption{Temperature dependence of magnetization measured with
cooling and warming cycles under the magnetic field of 5000 Oe for
the sample $BaCo_{0.9}Ni_{0.1}S_{1.97}$.\\ }
\end{figure}

Magnetic properties were systematically studied with cooling and
warming cycles as the second sequence in resistivity measurement.
Temperature dependence of magnetization is shown in Fig.2 under
magnetic field (H) of 5000 Oe. Figure 2 shows an antiferromagnetic
transition at $T_N \sim 270 K$ and a sharp change around $T_{s}$
on cooling from $300 K \rightarrow 5 K\rightarrow 220K$. The sharp
change around $T_{s}$ could arise from spin-sate change of Co ions
from high-spin to low-spin. Neutron diffraction data have shown
that the Co ions of $BaCoS_2$ have a localized high-spin
configuration with s=3/2\cite{kodama}. Pressure can induce a spin
state transition from the localized high-spin to low-spin
configuration\cite{zhou}. Therefore, the kink at $T_{s}$ observed
in resistivity is associated with spin-state change. Upon warming,
another sharp change of magnetization is observed at $T_{s1}\sim89
K$, which corresponds to the sharp change in resistivity measured
on warming in the first round of the second sequence as shown in
Fig.1. It indicates that the resistivity behavior is closely
related to the magnetic property. Being consistent with
resistivity results, the first-order transition at $\sim 140 K$ is
also observed in magnetization in the second round from $220 K
\rightarrow 5 K\rightarrow 220K$, but is gone in the continuous
measurement with cooling the sample from 400 K. It further
indicates that the "400 K" plays an "annealing" role and remove
memory effect. It should be pointed out that the intriguing
phenomena are directly related to the change in resistivity and
magnetization at $T_{s}$. Such change at $T_s$ can be observed
only in the sample $BaCo_{0.9}Ni_{0.1}S_{2-\delta}$ with narrow
sulphur deficiency around $\delta \sim 0.03$  close to the phase
boundary between an AFI at lower sulphur deficiency and a PMM at
higher sulphur deficiency.

\begin{figure}[h]
\includegraphics[width=9cm]{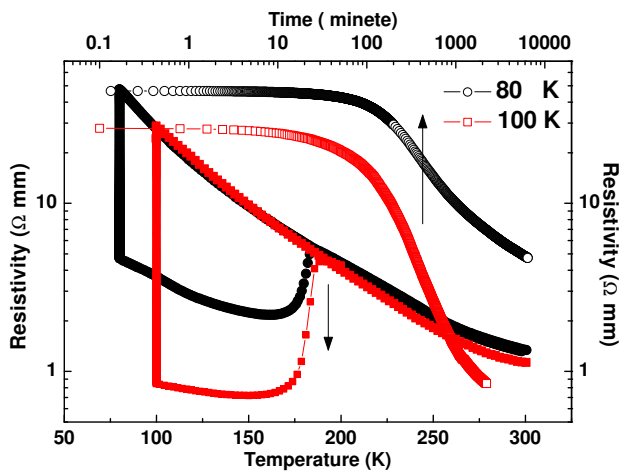}
\caption{(a): Temperature dependence of resistivity measured at
cooling rate of 3 K/min with an intermediate relaxation at 80 K
and 100 K for sample $BaCo_{0.9}Ni_{0.1}S_{1.97}$, respectively
(solid); (b): The same data at 80 K and 100 K are plotted as a
function of time (open).\\ }
\end{figure}

In order to understand the anomalous evolution of resistivity and
magnetization with the cooling and warming cycles, a systematic
study of the relaxation (i.e. time dependence) was carried out. As
shown in Fig.3 and Fig.4, the relaxation of resistivity and
magnetization at 80 K and 100 K was studied after the sample was
cooled from 300 K to the desired temperature with the same cooling
rate, respectively. It should be pointed out that the sample must
have the exact same experience before relaxation measurement
because the properties of the sample are strongly dependent on the
experience history and what temperature the sample was cooled from
as shown in Fig.1 and Fig.2. Therefore, the sample was warmed up
to room temperature and kept for a long time (more than ten days)
to make sure that the sample has the same behavior before cooling
to the next desired temperature after relaxation study at one
temperature. Fig.3 and Fig.4 show that the isothermal resistivity
and magnetization at 80 and 100 K decrease apparently with
relaxation time. Both relaxations are consistent with the
isothermal growth of PMM regions embedded in a AFI host. Such
nonstationary behavior has been described as "physical aging" and
is widely observed in the glass system.\cite{hodge} The plots of
resistivity and magnetization at 80 K and 100 K as a function of
time show a slow,
\begin{figure}
\includegraphics[width=9cm]{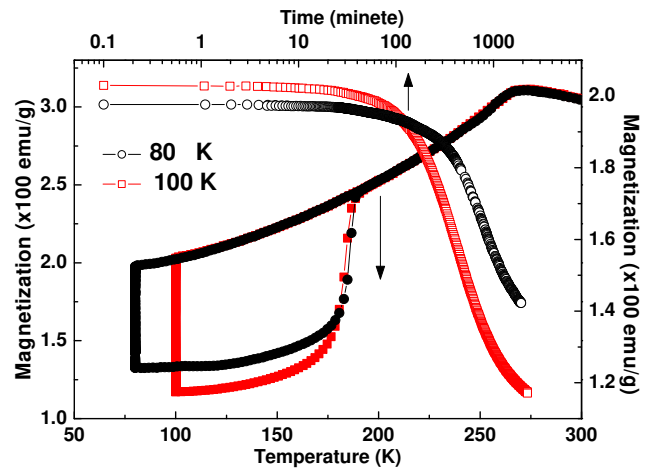}
\caption{(a): Temperature dependence of magnetization measured at
cooling rate of 3 K/min with an intermediate relaxation at 80 K
and 100 K under 5000 Oe for the sample
$BaCo_{0.9}Ni_{0.1}S_{1.97}$, respectively (solid); (b): The same
data at 80 K and 100
K are plotted as a function of time (open).\\
}
\end{figure}
nonexponential relaxation in both of resistivity and
magnetization, which is typical relaxation characteristic of glass
dynamics. Such glass behavior is not like that of a glass
consisting purely of spin or charge, but a cross-coupled variable.
In contrast to all relaxations observed so far, the resistivity
and magnetization nearly keep unchanged with a relaxation time of
about 100 minutes. After that, the resistivity and magnetization
decrease remarkably, but do not follow the exponential relaxation
behavior. The resistivity decreases by one and half order of
magnitude at 100 K with relaxation time of about 2000 minutes,
finally reaches nearly the same value as that in PMM phase as
shown in Fig.1. It suggests that the change in both resistivity
and magnetization with relaxation at fixed temperature arises from
a spontaneous phase transformation from the AFI with HTT structure
to PMM with LTM structure without external perturbation. This is
confirmed by x-ray diffraction measurements. X-ray diffraction
indicates that the structure of the sample changes with relaxation
at 100 K from tetragonal to monoclinic structure.

The relaxation behavior of resistivity and magnetization shown in
Fig.3 and Fig.4 is qualitatively similar to the nucleation process
of crystals\cite{mullin}. It further indicates that the striking
feature observed in Fig.1 and Fig.2 arises from the nucleation of
PMM in the AFI matrix. For the nucleation, a period of time
usually elapses between achievement of supersaturation or
supercooling and appearance of crystals. This time lag is
generally referred to an "induction period". Generally, the
physical properties do not change detectably in the induction
period. Therefore, the time spent before apparent change of
resistivity and magnetization shown in Fig.3 and Fig.4 can be
referred to the induction time. But it should be pointed out that
the induction time is normally very short (few
seconds)\cite{mullin}. However, the induction period in current
material is \emph{extremely long }(about 100 minutes). Which is
very \emph{unusual}, and may be the reason why the striking
behaviors shown in Fig.1 and Fig.2 occur.

The intriguing phenomena observed above can be understood by the
anomalous nucleation of the PMM phase with LTM in the AFI matrix
with HTT. The nucleation leads to enhancement of the PMM phase
with respect to the AFI phase, and the resistivity and
magnetization vary with the motion of the domain boundaries
separating the coexisting phases. The metastable state consisting
of PMM clusters and an AFI matrix results in a slow relaxation
dynamics of the resistivity and magnetization between $T_s$ and
$T_{c1}$. Theoretical work by Ahn et al.\cite{ahn} has indicated
that the combined effects of long-range strain field and local
intrinsic disorder naturally give rise to the phase separation and
a metastable landscape with hierarchical energy barriers for
relieving the strain, which can explain the phase separation in
manganites\cite{eblen} The phase separation occurred in
$BaCo_{0.9}Ni_{0.1}S_{1.97}$ should follow this mechanism. This is
confirmed with the suppression of AFI-PMM transition by
pressure\cite{looney}. Phase conversion within the admixture
involves rearrangement of many coupled degrees of freedom spanning
all relevant length scales. The presence of competing strain
fields, Coulomb interactions, magnetic correlations, and disorder
may \emph{frustrate the nucleation}, giving rise to the complex
free energy landscape with hierarchical energy barriers in
$BaCo_{0.9}Ni_{0.1}S_{1.97}$. This naturally give rise to glassy
dynamics, and a slow, nonexponential relaxation and "physical
aging" behavior\cite{palmer}.

Below $T_c$, the PMM regions with LTM structure start to grow
against the host material with the equilibrium size of the
clusters increasing as T is lowered. The dynamical process
followed by the clusters to reach their equilibrium size can be
thought of as a stepwise movement of the phase boundaries through
energy barriers. In the case, the existence of a hierarchy of
energy barriers is revealed by the response of the hysteresis in
resistivity and magnetization shown in Fig.1 and Fig.2 to the
cooling and warming cycles. The cooling in the cycles lowers the
free energy of PMM phase with respect to the AFI phase, and the
interphase domain walls that separate the PMM and AFI regions in
the mixed phase feel an effective force. Pinning sites up to a
certain strength are then overcome and effectively eliminated, and
the walls move irreversibly into a new configuration. This pinning
behavior arises from the competition of strain fields, Coulomb
interactions, magnetic correlations, and disorder in strongly
correlated electronic system. These pinning sites can completely
be removed only at $\sim \emph{400 K}$. This is the reason why
effect of \emph{400 K} on resistivity and magnetization occurs.
Such dynamical process produces aging and memory effects in
resistivity and magnetization. Such anomalous relaxation observed
here without external perturbation is quite different from the
normal relaxation which can be only observed after driving the
system out-of-equilibrium with external field.

In conclusion, we found \emph{an unusual, and extremely slow and
nonexponential} relaxation of resistivity and magnetization in
$BaCo_{0.9}Ni_{0.1}S_{1.97}$. Another \emph{unusual} feature is
that the induction time for the PMM nucleation in AFI matrix is
extremely long in contrast to normal nucleation. All intriguing
phenomena arise from the strong cross-couplings between the
different degrees of freedom and competition of strain fields,
Coulomb interactions, magnetic correlations and disorders, leading
to complicated and slow relaxation. Similar phenomena should be
expected in other strongly correlated electronic system in phase
separation region around MIT.

{\bf Acknowledgment:} We would like to thank Drs. X. G. Wen, D. L.
Feng and S. Y. Li for useful discussion. This work is supported by
the Nature Science Foundation of China and by the Ministry of
Science and Technology of China (973 project No: 2006CB601001) and
by National Basic Research Program of China (2006CB922005).

\end{document}